\documentclass{sobolev}    
\setpg{1}                  

\titl[Search for the Giant Pulses]{Search for the Giant Pulses \\ - an extreme phenomenon in radio pulsar emission}

\authors{A.N.\,Kazantsev, V.A.\,Potapov}{A.N.\,Kazantsev\aff{1,2}, V.A.\,Potapov\aff{1}}

\affiliat{
 \item P.N.Lebedev Physical Institute of the RAS, Pushchino Radio Astronomy Observatory ASC LPI RAS, Russia   
 \item Pushchino State Institute of Natural Sciences, Russia
}

\email{kaz.prao@bk.ru}   

\usepackage{amsmath}
\usepackage{amssymb}
\usepackage{indentfirst}

\begin{document} 

\begin{abstract}
  
Here we present results of our search for Giant Pulses(GPs) from pulsars of Northern Hemisphere. Our survey was carried out at a frequency of 111 MHz using the Large Phased Array (LPA) radio telescope. Up to now we have detected regular generation of strong pulses satisfying the criteria of GPs from 2 pulsars: B1133+16, B1237+25. 
  
\end{abstract}

\section{Introduction}\label{estimation}
Slow pulse-to-pulse variation of intensity is very typical for overwhelming majority of known pulsars. However, a handful of pulsars have mysterious mechanism which can break such a stability.

Phenomenon of generation GPs was first detected for Crab pulsar(B0531+21)~\cite{Staelin} and the millisecond pulsar B1937+21~\cite{Wolszczan}. In following studies a set of typical characteristics of GPs was determined. These are: very narrow components in pulse's microstructure (up to several nanoseconds), high peak flux density(up to several MJy), and power-law distribution of the peak flux density of GPs. It is important to emphasize, that the pulsar B0531+21 and B1937+21 have very strong magnetic field on the light cylinder ($B_{LC} \approx 10^6~G$) and are considered as "classical" pulsars with GPs. In course of time, GPs were detected from 5 pulsars with very strong magnetic field on the light cylinder ($B_{LC} \approx 10^5-10^6 ~G$): B0218+42~\cite{Joshi}, B0540-69~\cite{Johnston2003}, B1821-24~\cite{Johnston2001}, J1823-3021~\cite{Knight} and B1957+20~\cite{Joshi}.

However, later on similar phenomenon was detected (mostly at low radio frequencies between 40 and 111 MHz) for set of pulsars with a value of $B_{LC}$ from several to several hundreds gauss: J0034-0721~\cite{Kuzmin2004}, J0529-6652~\cite{Crawford}, J0659+1414~\cite{Kuzmin2006}, J0953+0755~\cite{Singal, Smirnova}, J1115+5030~\cite{Ershov2003} and J1752+2359~\cite{Ershov2006}. 

In present work we describe the results of our observations that were held to search for new pulsars generating GPs or anomalous strong pulses at low radio frequencies.

\section{Observations and Processing}
The observations were made during 2012 and 2014 at the Pushchino Radio Astronomy Observatory with the Large Phased Array radio telescope. This is the transit telescope with the effective area of about $20000 \pm 4000 ~m^2$ in the zenith direction. The main frequency of the observations was 111~MHz with a bandwidth of 2.3 MHz(460$\times$5 kHz channels digital receiver with post-detector DM (Dispersion Measure) removal). The sampling interval was 1.2288 ms and the duration of each observation session was about 3.5 min (153 periods of PSR B1237+25).

We have processed the results of 66 observational sessions containing 11 091 pulses of B1133+16 and 89 observational sessions containing 13 617 pulses of B1237+25.

Average pulse of pulsar was obtained by summing and averaging of all individual pulses during one session of pulsar's observation. We have analyzed every pulse with a peak flux density $ > 4 \sigma_{noise}$ and located at the phase of the average pulse. Pulses with peak flux density  more than 30 peak flux density of the average (per session) pulse were marked as GPs candidates.

For PSR B1133+16 we calculated peak flux density distribution separately for two main components of the average pulse of pulsar.

\section{Results and Conclusion}
B1133+16 and B1237+25 are active second period (normal) radio pulsars with multicomponent average pulses having two and five main components respectively. B1133+16 has a rotational period $P=1.1879~s $ and $B_{LC}=11.9~G$, B1237+25 --  $P=1.3824~s$ and $B_{LC}=4.14~G$.  We have regularly observed strong pulses from each pulsar during entire period of our observation ~\cite{Kaz_Pot1, Kaz_Pot2}.

An example of GPs from B1237+25 and B1133+16 are shown in Fifure 1. The most powerful GP of B1237+25 was detected on August 12, 2012, and have a flux density of 900$\pm$130~Jy.  There are around 12 GP events per 10000 pulses for B1237+25, and around 16 GP events per 10000 pulses for B1133+16.

The distribution of the peak flux density (in signal to noise ratio units) in the Log-Log scale is shown in Figure 2. For B1237+25 distribution of the strong pulses has a bimodal power-law shape with exponents -1.26$\pm$0.05 and -3.36$\pm$0.34, which is quite typical of GPs and obviously differs from log-normal individual pulses distribution of regular pulses. The distribution for B1133+16 has a complex character and may be fitted as a combination of two log-normal distributions for the first component of pulse and two log-normal and one power-law components with an exponent -2.39 $\pm$ 0.08, for the second.

\begin{figure}[th]
  \centering
   \includegraphics[width=.75\paperwidth]{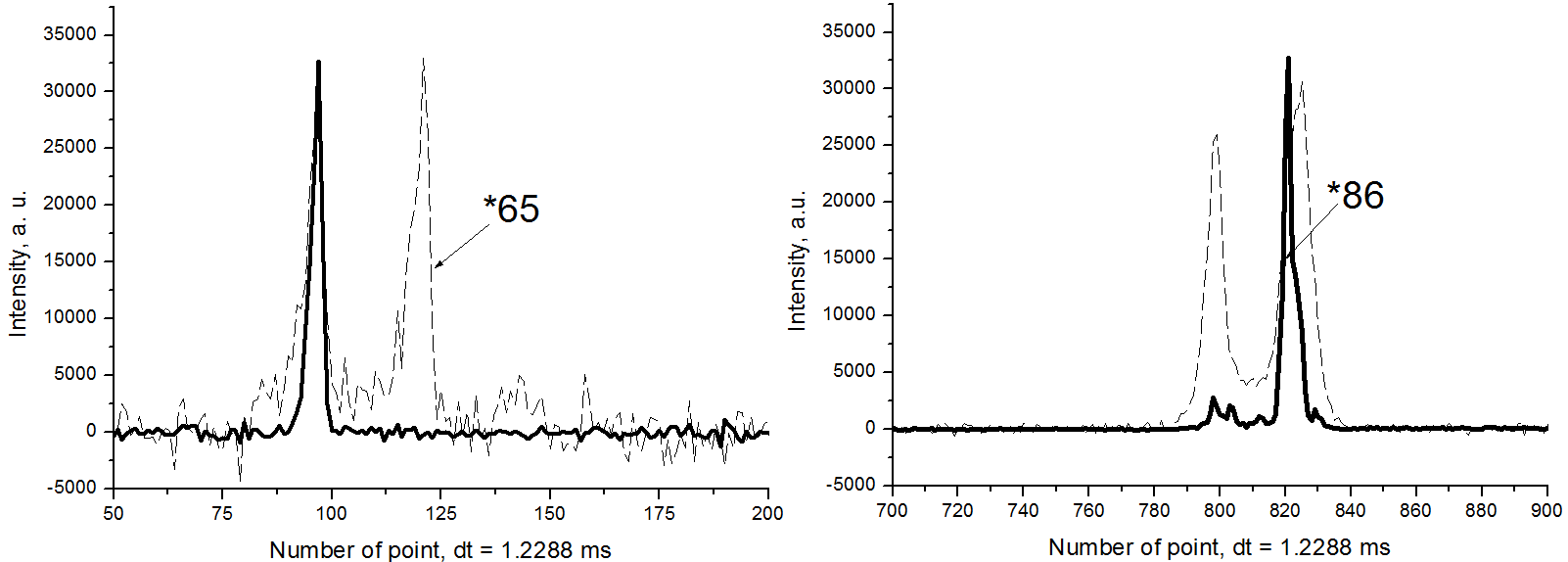}
  \caption{Strong pulses of PSR B1237+25 observed on May 14, 2012 (left panel), and of PSR B1133+16 observed on March 14, 2014 (right panel). The average pulse (dashed line) is shown multiplied by 65 and 86, respectively.}
  \label{pulses}
\end{figure}

\begin{figure}[th]
  \centering
  \includegraphics[width=0.75\paperwidth]{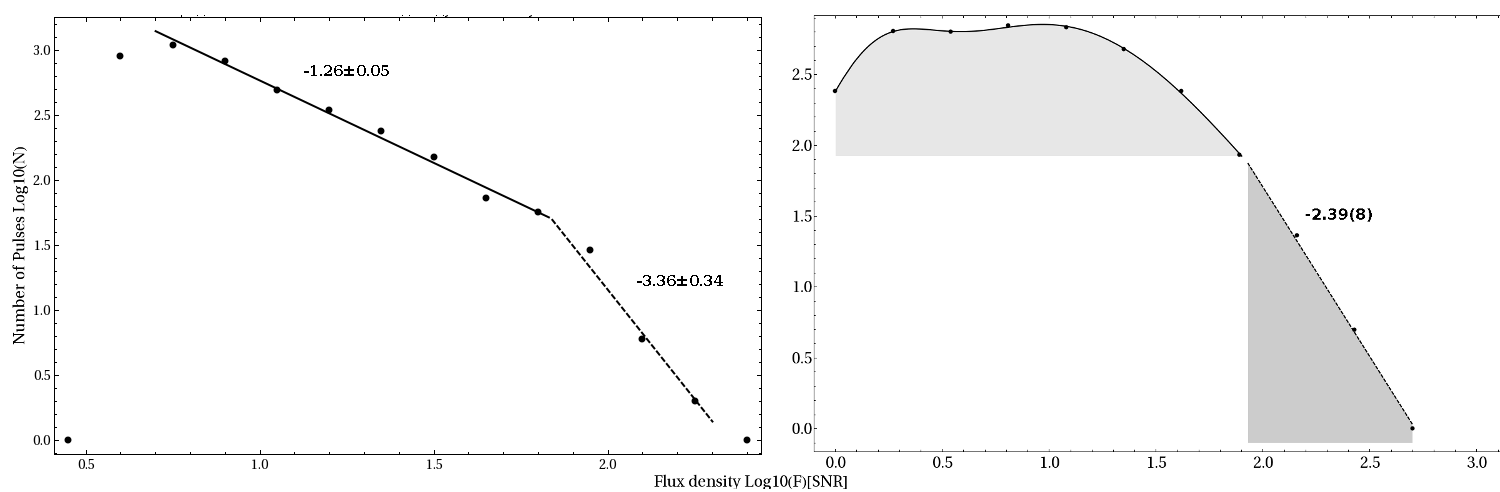}
  \caption{Histograms of distributions of the peak flux density of individual pulses for B1237+25 (left panel), and for the second (right-hand) component of B1133+16 (right panel) shown in the Log-Log scale. Data was fitted by the combination of two power-law distributions, and two log-normal plus power-law distribution, respectively.}
  \label{distrib}
\end{figure}

We can conclude that strong individual pulses of PSR B1133+16 and PSR B1237+25 observed at 111~MHz satisfy the main criteria of GPs. It worth noting that B1133+16 and PSR B1237+25 are pulsars with low magnetic field on light cylinder. This is the further confirmation of our earlier assumptions that such pulsars (including J0034-0721, J0529-6652, J0659+1414, J0953+0755, J1115+5030 and J1752+2359) may be referred as a sub-class of  pulsars with GP, having low magnetic field on light cylinder and generating GPs mostly at low radio frequencies.

\section{Acknowledgements}
This work was supported by the Program of the Presidium of the Russian Academy of Sciences 41P "Transitional and explosive processes in astrophysics", and by Scientific Educational Complex Program of P.N. Lebedev Institute of the RAS.


\begin{thebibliography}{999}

\bibitem{Staelin}
\textit{D.N.\,Staelin, E.C.\,Reifenstein}, IAU Astron. Telegram Circ, \textbf{2110}, 1968.

\bibitem{Wolszczan} 
\textit{A.\,Wolszczan, J.\,Cordes, D.\,Stinebring}, NRAO Workshop, 63, 1984.

\bibitem{Joshi}
\textit{B.C.\,Joshi, M.\,Kramer, A.G.\,Lyne, M.\,McLaughlin, I.H.\,Stairs}, IAUS, \textbf{218}, 319, 2004.

\bibitem{Johnston2003}
\textit{S.\,Johnston, R.W.\,Romani}, ApJL, \textbf{590}, 95, 2003.

\bibitem{Johnston2001}
\textit{S.\,Johnston,  R.W.\,Romani}, ApJL, \textbf{557}, 93, 2001.

\bibitem{Knight}
\textit{H.S.\,Knight, M.\,Bailes, R.N.\,Manchester, S.M.\,Ord}, ApJ, \textbf{625}, 951, 2005.

\bibitem{Kuzmin2004}
\textit{A.D.\,Kuzmin, A.A.\,Ershov, B.Ya.\,Losovsky}, Astron. Lett., \textbf{30}, 247, 2004.

\bibitem{Crawford}
\textit{F.\,Crawford, D.\,Altemose, H.\,Li, D.R.\,Lorimer}, ApJ, \textbf{762}, 97, 2013.

\bibitem{Kuzmin2006}
\textit{A.D.\,Kuzmin, A.A.\,Ershov}, Astron. Lett., \textbf{32}, 583, 2006.

\bibitem{Singal}
\textit{A.K.\,Singal}, Astrophys. and Space Sci., \textbf{278}, 61, 2001.

\bibitem{Smirnova}
\textit{T.V.\,Smirnova}, Astron. Rep., \textbf{54}, 430, 2012.

\bibitem{Ershov2003}
\textit{A.A.\,Ershov, A.D.\,Kuzmin}, Astron. Lett., \textbf{29}, 91, 2003.

\bibitem{Ershov2006}
\textit{A.A.\,Ershov, A.D.\,Kuzmin}, Chin. J. Astron. and Astrophys., \textbf{6}, 30, 2006.

\bibitem{Kaz_Pot1}
\textit{A.N.\,Kazantsev, V.A.\,Potapov}, Astronomicheskii Tsirkulyar, \textbf{1620}, 1-7, 2015.

\bibitem{Kaz_Pot2}
\textit{A.N.\,Kazantsev, V.A.\,Potapov}, Astronomicheskii Tsirkulyar, \textbf{1628}, 1-8, 2015.

\end{thebibliography}
\end{document}